\documentclass[aps,showpacs,onecolumn,11pt]{revtex4}
\usepackage{epsfig,colordvi}
\usepackage{graphics}
\usepackage{amssymb,amsmath}
\usepackage{times}
\newcommand{\ba}{\begin{eqnarray}}
\newcommand{\ea}{\end{eqnarray}}
\newcommand{\bsub}{\begin{subequations}}
\newcommand{\esub}{\end{subequations}}
\newcommand{\tl}{\tilde{\ell}}
\newcommand{\tL}{\tilde{L}}

\begin{document}

\sloppy \raggedbottom

 \setcounter{page}{1}

\title{Partial Dynamical Symmetry in Odd-Mass Nuclei}

\author{A. Leviatan}{}

\affiliation{Racah Institute of Physics, The Hebrew University,
Jerusalem 91904, Israel}{}

\begin{abstract}
Spectral features of the odd-mass nucleus $^{195}$Pt are analyzed
by means of an interacting boson-fermion Hamiltonian with
SO(6) partial dynamical symmetry. For the latter,
selected eigenstates are solvable and preserve the symmetry exactly,
while other states are mixed. The analysis constitutes a first example
of this novel symmetry construction in a mixed Bose-Fermi system.
\end{abstract}
\maketitle

\section[]{Introduction}

The concept of dynamical symmetry (DS) has been widely used
to interpret nuclear structure.
A given DS admits an analytic solution
for all states of the system, with characteristic degeneracies,
quantum numbers and selection rules.
Familiar examples are the U(5), SU(3)
and O(6) DSs of the interacting boson model (IBM~\cite{ibm}) 
of even-even nuclei, 
which encode the dynamics of spherical, axially-deformed and
$\gamma$-unstable nuclear shapes.
The majority of nuclei, however, exhibit strong
deviations from these solvable benchmarks.
More often one finds that the assumed symmetry is not obeyed uniformly,
{\it i.e.}, is fulfilled by some of the states but not by others.
The need to break the DSs, but still preserve important symmetry
remnants, has led to the introduction of partial
dynamical symmetry (PDS)~\cite{Leviatan11}.
For the latter case, only selected eigenstates of the Hamiltonian
retain solvability and good symmetry,
while other states are mixed.
Various types of PDSs were proposed and algorithms for constructing
Hamiltonians with such property have been
developed~\cite{Leviatan11,Alhassid92,GarciaRamos09}.
Bosonic Hamiltonians with PDS
have been applied to nuclear spectroscopy,
where extensive tests provide empirical evidence for
their relevance to a broad range of
nuclei~\cite{Leviatan96,lev99,Casten14,Couture15,Leviatan13,
GarciaRamos09,Leviatan02,Kremer14}.
Fermionic shell model Hamiltonians with PDS
have been applied to light nuclei~\cite{Escher00,Escher02}
and seniority isomers~\cite{Rowe01,Rosen03,Isacker08,isa14}.
These empirical manifestations and further applications to
nuclear shape-phase transitions~\cite{Leviatan07,Macek14},
suggest a more pervasive role of PDSs in
nuclei than heretofore realized.

All examples of PDS considered so far,
were confined to systems of a given statistics.
In the present contribution,
we consider an extension 
of the PDS concept to mixed systems of bosons and fermions~\cite{isa15},
of relevance to odd-mass nuclei. 
As an example of such novel
symmetry construction, spectral features of $^{195}$Pt are analyzed
in the framework of the interacting boson fermion model.

\section{SO$^{\rm\bf BF}$(6) Dynamical Symmetry Limit of the IBFM}

The interacting boson fermion model (IBFM~\cite{ibfm}) describes properties
of low-lying states in odd-mass nuclei, in terms of
$N$ bosons ($b^{\dag}_{\ell,m}$) with $\ell=0$ ($s^{\dag}$)
and $\ell=2$ ($d^{\dag}_m$), representing valence nucleon pairs,
and a single fermion ($a^{\dag}_{j,m}$) in a shell model orbit 
with angular momentum $j$.
In the current study, $j=1/2,3/2,5/2$, which 
can be divided into a pseudo-orbital angular momentum $(\tilde\ell\!=\!0,2)$
coupled to a pseudo-spin ($\tilde s\!=\!1/2$). 
The $\tilde\ell$-$\tilde s$ and $jm$ bases are related by 
$c^{\dag}_{\tilde\ell\tilde m_\ell;\tilde s\tilde m_s} \!=\!
\sum_{j,m}(\tilde\ell,\tilde m_{\ell};\tilde s,\tilde m_s\vert j,m)
\,a^{\dagger}_{jm}$. 
The bilinear combinations $\{b^{\dag}_{\ell,m}b_{\ell',m'}\}$ 
and $\{a^{\dag}_{j,m}a_{j',m'}\}$ 
span, respectively, bosonic (B) and fermionic (F) algebras, 
forming a spectrum generating algebra 
${\rm U}^{\rm B}(6)\otimes{\rm U}^{\rm F}(12)$. 
The IBFM Hamiltonian is
expanded in terms of these generators and consists of Hermitian
rotational-invariant interactions which conserve the total number
of bosons,
$\hat{N}=s^{\dag}s + \sum_{m}d^{\dag}_md_m$, and of fermions
$\hat{n} = \sum_{j,m}a^{\dag}_{j,m}a_{j,m}$.

There exist several strategies to define
DSs with ${\rm U}^{\rm B}(6)\otimes{\rm U}^{\rm F}(12)$
as a starting point~\cite{ibfm}.
They all define a chain of nested subalgebras,
relying on the existence of isomorphisms between boson and fermion algebras
and ending in the symmetry algebra. 
Here we focus on the
${\rm SO}^{\rm BF}(6)$ DS limit of the model,
corresponding to the classification:
\begin{equation}
\begin{array}{l}
{\rm U}^{\rm B}(6)\otimes{\rm U}^{\rm F}(12)\\
\quad\downarrow\qquad\qquad\downarrow\\
\;\;[N]\qquad\quad\,[1]\\[2mm]
\supset
\Big({\rm U}^{\rm BF}(6)\supset{\rm SO}^{\rm BF}(6)
\supset{\rm SO}^{\rm BF}(5)\supset{\rm SO}^{\rm BF}(3)\Big)
\otimes{\rm SU}^{\rm F}(2)\supset{\rm Spin}^{\rm BF}(3)\\
\quad\qquad\downarrow\;\;\qquad\qquad\downarrow
\;\;\qquad\qquad\downarrow\;\;\qquad\qquad\downarrow
\;\;\,\qquad\qquad\downarrow\;\;\;\qquad\qquad\downarrow\\
\;\;\;\;\; [N_1,N_2]\;\;\;\;\;\;\;\langle\sigma_1,\sigma_2\rangle
\qquad (\tau_1,\tau_2)\;\;\;\,\qquad L
\qquad\qquad\;\;\tilde{s}\qquad\qquad\;\;\, J
\end{array}
\label{e_clas}
\end{equation}
where underneath each algebra ($G$)
the associated labels of the irreducible representations (irreps) 
are indicated.
The indicated Bose-Fermi algebra $G^{\rm BF}$ is the direct sum 
of $G^{\rm B}$ and $G^{\rm F}$.
\begin{table}[t]
\caption{\label{TabIBMcas}
\small
Generators and Casimir operators, $\hat{C}_{k}[G]$, of order
$k \!=\! 1,2$, for the Bose-Fermi algebras in Eq.~(\ref{e_hamds}). 
The generators are sums of bosonic: 
$(b^{\dag}_{\ell}\tilde{b}_{\ell'})^{(L)}$, and fermionic operators:
$K^{(\tL)}_{m}(\tl,\tl') \!=\! \sqrt{2}
(c^{\dag}_{\tl,1/2}\,\tilde{c}_{\tl',1/2})^{(\tilde{L},0)}_{m,0}$,
where $\tilde b_{\ell m_\ell}\!\equiv\!(-)^{\ell+m_\ell}b_{\ell,-m_\ell}$
and $\tilde a_{jm_j}\!\equiv\!(-)^{j+m_j}a_{j,-m_j}$. 
Here, $\hat{M}_{0} \!=\! s^{\dag}s + K^{(0)}_{m}(0,0)$, 
$\Pi^{(2)}_{m} \!=\! d^{\dag}_{m}s + s^{\dag}\tilde{d}_{m}
+ K^{(2)}_{m}(2,0)+K^{(2)}_{m}(0,2)$,
$\bar{\Pi}^{(2)}_{m} \!=\! i[d^{\dag}_{m}s - s^{\dag}\tilde{d}_{m}
+ K^{(2)}_{m}(2,0)- K^{(2)}_{m}(0,2)]$,
$U^{(\rho)}_{m} \!=\! (d^{\dag}\,\tilde{d})^{(\rho)}_{m} + K^{(\rho)}_{m}(2,2)$,
$\hat{L}_{m} \!=\! \sqrt{10}[U^{(1)}_{m} +K^{(1)}_{m}(2,2)]$ 
and $\hat{J}_{m} \!=\! \hat{L}_{m} + \hat{S}_{m}$, 
where 
$\hat{S}_{m} \!=\! \sum_{\tl=0,2}\sqrt{2\tl+1}
(c^{\dag}_{\tl,1/2}\,\tilde{c}_{\tl,1/2})^{(0,1)}_{0,m}$ 
are the pseudo-spin generators of SU$^{F}$(2).
The boson- and fermion number operators are 
$\hat{N} \!=\! s^{\dag}s + \sqrt{5}(d^{\dag}\tilde{d})^{(0)}$
and $\hat{n} \!=\; K^{(0)}_{m}(0,0) + \sqrt{5}K^{(0)}_{m}(2,2)$, 
respectively.}\smallskip
\begin{small}
\centering
\begin{tabular*}{\textwidth}{@{\extracolsep{\fill}}ll}
\hline  \noalign {\smallskip}
Algebra & Generators and Casimir operators $\hat{C}_{k}[G]$ \\
\hline  \noalign {\smallskip}
${\rm U}^{\rm BF}(6)$ & $\hat{M}_0,\, \Pi^{(2)},\, \bar{\Pi}^{(2)},\,
{\rm U}^{(\rho)}\; \rho = 0\!-\!4$;
$\quad \hat C_1[{\rm U}^{\rm BF}(6)] = \hat{N}+\hat{n}$\\[1mm]
& $\hat C_2[{\rm U}^{\rm BF}(6)] = \hat{M}^{2}_0 +
{\textstyle\tfrac{1}{2}\Pi^{(2)}\cdot\Pi^{(2)}
+\tfrac{1}{2}}\bar{\Pi}^{(2)}\cdot\bar{\Pi}^{(2)}
+\sum_{\rho=0\!-\!4}U^{(\rho)}\cdot U^{(\rho)}\qquad\qquad$\\[1mm]
${\rm SO}^{\rm BF}(6)$ &
$\Pi^{(2)},\, {\rm U}^{(1)},\,{\rm U}^{(3)}$;
$\quad\hat C_2[{\rm SO}^{\rm BF}(6)]=
2\sum_{\rho=1,3}U^{(\rho)}\cdot U^{(\rho)}
+ \Pi^{(2)}\cdot\Pi^{(2)}$ \\[1mm]
${\rm SO}^{\rm BF}(5)$ &
${\rm U}^{(1)},\,{\rm U}^{(3)}$;
$\quad\hat C_2[{\rm SO}^{\rm BF}(5)]=
2\sum_{\rho=1,3}U^{(\rho)}\cdot U^{(\rho)}$\\[1mm]
${\rm SO}^{\rm BF}(3)$ &
$\hat{L}_m$;
$\quad\hat C_2[{\rm SO}^{\rm BF}(3)] = \hat{L}\cdot \hat{L}$\\[1mm]
${\rm Spin}^{\rm BF}(3)$ &
$\hat{J}_m$;
$\quad\hat C_2[{\rm Spin}^{\rm BF}(3)] = \hat{J}\cdot\hat{J}$\\[1mm]
\hline  \noalign {\smallskip}
\end{tabular*}
\label{Tab1}
\end{small}
\end{table}

The eigenstates~(\ref{e_clas}) are obtained with a Hamiltonian
that is a combination of Casimir operators $\hat C_k[G]$
of order $k$ of an algebra $G$ appearing in the chain.
Up to a constant energy, this Hamiltonian is of the form
\begin{eqnarray}
\hat H_{\rm DS}&=&
a\,\hat C_2[{\rm U}^{\rm BF}(6)]+
b\,\hat C_2[{\rm SO}^{\rm BF}(6)]+
c\,\hat C_2[{\rm SO}^{\rm BF}(5)]
\nonumber\\&&+
d\,\hat C_2[{\rm SO}^{\rm BF}(3)]+
d'\hat C_2[{\rm Spin}^{\rm BF}(3)] ~.
\label{e_hamds}
\end{eqnarray}
Explicit expressions for the above Casimir operators are given in
Table~\ref{Tab1}.
The associated eigenvalue problem is analytically solvable,
leading to the energy expression
\ba
E_{\rm DS} &=&
a\,[N_1(N_1+5)+N_2(N_2+3)]
+ b\,[\sigma_1(\sigma_1+4)+\sigma_2(\sigma_2+2)] \qquad
\nonumber\\
&&
+ c\,[\tau_1(\tau_1+3)+\tau_2(\tau_2+1)]
+ d\,L(L+1) + d'J(J+1) ~.
\ea
The energy spectrum of the Hamiltonian~(\ref{e_hamds})
is then determined once the allowed values of
$[N_1,N_2]$, $\langle\sigma_1,\sigma_2\rangle$, $(\tau_1,\tau_2)$, $L$,
and $J$ for a given $N$ are found.
Such branching rules can be obtained
with standard group-theoretical techniques~\cite{ibfm}. 
The lowest-lying states in the spectrum of an
odd-mass nucleus, described in terms of $N$ bosons and one fermion,
can be classified as $|[N+1]\langle N+1\rangle(\tau)LJM_J\rangle$
with $\tau=0,1,\ldots N+1$.
The next class of states belongs to
$|[N,1]\langle N,1\rangle(\tau_1,\tau_2)LJM_J\rangle$
with $(\tau_1,\tau_2)=(\tau,0)\,{\rm or}\,(\tau,1)$
and $\tau=1,2,,...N$. 
There is also some evidence from one-neutron transfer
for $|[N,1]\langle N-1\rangle(\tau)LJM_J\rangle$ states~\cite{Metz00},
with $\tau=0,1,\ldots N-1$. The $L$-values of these states are obtained
from the known $SO(5)\supset SO(3)$ branching rules~\cite{ibfm}
and $J=L\pm 1/2$.

\section{SO$^{\rm\bf BF}$(6) Partial Dynamical Symmetry in the IBFM}

While $\hat{H}_{\rm DS}$~(\ref{e_hamds}) is {\em completely} solvable,
the question arises whether terms can be added
that preserve solvability for {\em part} of its spectrum.
This can be achieved by the construction of a PDS.
\begin{table}[t]
\caption{\label{t_tensors}
\small
Two-particle tensor operators in the ${\rm SO}^{\rm BF}(6)$ limit.
For the Bose-Fermi pairs, the superscript ${\cal L}({\cal J})$ stands for 
the coupling ${\cal J}={\cal L}\pm 1/2$.}\smallskip
\begin{small}
\centering
\begin{tabular*}{\textwidth}{@{\extracolsep{\fill}}l}
\hline  \noalign {\smallskip}
$B^{\dag}_{[N_1,N_2]\langle\sigma_1,\sigma_2\rangle(\tau_1,\tau_2){\cal L}
({\cal J})}\equiv{\cal T}^{{\cal L}({\cal J})}_{+,M_{\cal J}}$\\[2mm]
\hline  \noalign {\smallskip}
$B^{\dag}_{[2,0]\langle0,0\rangle(0,0)0(0)}\;\;\, \equiv
{\cal V}^{0(0)}_+ \;\;\; =
{\textstyle\sqrt{\frac5{12}}(d^\dag d^\dag)^{(0)}_0
-\sqrt{\frac1{12}}(s^\dag s^\dag)^{(0)}_0}$\\[1mm]
$B^{\dag}_{[2,0]\langle0,0\rangle(0,0)0(1/2)} \equiv
{\cal V}^{0(1/2)}_{+,\mu} =
{\textstyle -\sqrt{\frac16}(s^\dag a^\dag_{1/2})^{(1/2)}_\mu-
\sqrt{\frac13}(d^\dag a^\dag_{3/2})^{(1/2)}_\mu}
{\textstyle + \sqrt{\frac12}(d^\dag a^\dag_{5/2})^{(1/2)}_\mu}$\\[2mm]
$B^{\dag}_{[1,1]\langle1,1\rangle(1,1)1(1/2)} \equiv
{\cal U}^{1(1/2)}_{+,\mu} =
{\textstyle\sqrt{\frac35}(d^\dag a^\dag_{3/2})^{(1/2)}_\mu+
\sqrt{\frac25}(d^\dag a^\dag_{5/2})^{(1/2)}_\mu}$\\[1mm]
$B^{\dag}_{[1,1]\langle1,1\rangle(1,1)1(3/2)} \equiv
{\cal U}^{1(3/2)}_{+,\mu} =
{\textstyle -\sqrt{\frac{3}{10}}(d^\dag a^\dag_{3/2})^{(3/2)}_\mu+
\sqrt{\frac{7}{10}}(d^\dag a^\dag_{5/2})^{(3/2)}_\mu}$\\[2mm]
$B^{\dag}_{[1,1]\langle1,1\rangle(1,1)2(3/2)} \equiv
{\cal U}^{2(3/2)}_{+,\mu} =
\textstyle{\sqrt{\frac12}(s^\dag a^\dag_{3/2})^{(3/2)}_\mu-
\sqrt{\frac12}(d^\dag a^\dag_{1/2})^{(3/2)}_\mu}$\\[1mm]
$B^{\dag}_{[1,1]\langle1,1\rangle(1,1)2(5/2)} \equiv
{\textstyle{\cal U}^{2(5/2)}_{+,\mu} =
\sqrt{\frac12}(s^\dag a^\dag_{5/2})^{(5/2)}_\mu-
\sqrt{\frac12}(d^\dag a^\dag_{1/2})^{(5/2)}_\mu}$\\[1mm]
$B^{\dag}_{[1,1]\langle1,1\rangle(1,1)3(5/2)} \equiv
{\cal U}^{3(5/2)}_{+,\mu} =
{\textstyle\sqrt{\frac45}(d^\dag a^\dag_{3/2})^{(5/2)}_\mu+
\sqrt{\frac15}(d^\dag a^\dag_{5/2})^{(5/2)}_\mu}$\\[1mm]
$B^{\dag}_{[1,1]\langle1,1\rangle(1,1)2(7/2)} \equiv
\textstyle{{\cal U}^{3(7/2)}_{+,\mu} =
-\sqrt{\frac{1}{10}}(d^\dag a^\dag_{3/2})^{(7/2)}_\mu+
\sqrt{\frac{9}{10}}(d^\dag a^\dag_{5/2})^{(7/2)}_\mu}$\\[2mm]
\hline  \noalign {\smallskip}
\end{tabular*}
\label{Tab2}
\end{small}
\end{table}

The algorithm to construct an Hamiltonian with a PDS
is based on its expansion, $\hat{H}' = \sum_{\alpha,\beta}
u_{\alpha\beta}\, \hat{B}^{\dag}_{\alpha}\hat{B}_{\beta}$,
in terms of tensors
which annihilate prescribed set of states~\cite{Alhassid92,GarciaRamos09}. 
The tensors of interest in the present study, are listed in
Table~\ref{t_tensors}. They are composed of two-particle operators 
(either two bosons or a boson-fermion pair), 
and have definite character under the chain~(\ref{e_clas}),
$B^{\dag}_{[N_1,N_2]\langle\sigma_1,\sigma_2\rangle(\tau_1,\tau_2){\cal L}
({\cal J})}\equiv{\cal T}^{{\cal L}({\cal J})}_{+,M_{\cal J}}$. 
The corresponding annihilation operators with the correct tensor
properties follow from
$\tilde{\cal T}^{{\cal L}({\cal J})}_{-,M_{\cal J}} \!=\!
(-)^{{\cal J}+M_{\cal J}}\left({\cal T}^{{\cal L}({\cal J})}_{+,-M_{\cal J}}\right)^\dag$,
where ${\cal T}\!=\!{\cal U}$ or $\cal V$.
All these operators 
annihilate particular states, hence lead to a PDS of some kind.
For example, the operators
with ${\rm U}^{\rm BF}(6)$ labels $[N_1,N_2]=[1,1]$ satisfy
\begin{equation}
\tilde{\cal U}^{{\cal L}({\cal J})}_{-,M_{\cal J}}
|[N+1]\langle\sigma\rangle(\tau)LJM_J\rangle=0 ~,
\label{e_anni1}
\end{equation}
for all permissible $(\sigma \tau L J M_J)$.
This is so because a state with $N-1$ bosons and no fermion
has the ${\rm U}^{\rm BF}(6)$ label $[N-1]$.
Given the multiplication rule $[N-1]\otimes[1,1]=[N,1]\oplus[N-1,1,1]$,
the action of a ${\cal U}^{{\cal L}({\cal J})}_{+,-M_{\cal J}}$ operator
on an $(N-1)$-boson state can never yield a boson-fermion state
with the ${\rm U}^{\rm BF}(6)$ labels $[N+1]$.
Similar arguments involving SO(6) multiplication
lead to the following properties for the $\cal V$ operators
which have SO(6) tensor character $\langle0,0\rangle$:
\begin{subequations}
\begin{eqnarray}
&&\tilde{\cal V}^{0({\cal J})}_{-,M_{\cal J}}
|[N+1]\langle N+1\rangle(\tau)LJM_J\rangle=0\;,\\
&&\tilde{\cal V}^{0({\cal J})}_{-,M_{\cal J}}
|[N,1]\langle N,1\rangle(\tau_1,\tau_2)LJM_J\rangle=0\;.
\end{eqnarray}
\label{e_anni2}
\end{subequations}
Normal-ordered interactions 
with PDS can now be constructed out of the $\cal T$-operators in
Table~\ref{t_tensors}, as
\ba
\hat{H}' &=&
x_{00}^{0}\,({\cal V}^{0({0})}_+\tilde{\cal V}^{0(0)}_-)^{(0)}
+x_{00}^{1/2}\,\sqrt{2}\,({\cal V}^{0({1/2})}_+\tilde{\cal V}^{0(1/2)}_-)^{(0)}
\nonumber\\
&&
+ \sum_{\cal LL'J}x_{\cal LL'}^{\cal J}\,\sqrt{2{\cal J}\!+\!1}\,
[({\cal U}^{\cal L(J)}
\tilde{\cal U}^{\cal L'(J)}_-)^{(0)} +{\rm H.c.}]
\nonumber\\
&&
+\, x_{10}^{1/2}\,\sqrt{2}\,[
({\cal U}^{1(1/2)}_+\tilde{\cal V}^{0(1/2)}_-)^{(0)}
+{\rm H.c.}] ~,
\label{Hprime}
\ea
where H.c. stands for Hermitian conjugate. 
Particular linear combinations of terms in Eq.~(\ref{Hprime}) yield the 
Casimir operators in $\hat{H}_{\rm DS}$, Eq.~(\ref{e_hamds}).
Specifically, the quadratic Casimir operator of ${\rm U}^{\rm BF}(6)$ 
is obtained for
\ba
\hat{{\cal M}}_{6} &\equiv&
(\hat{N} + \hat{n})(\hat{N} + \hat{n} +5)
-\hat C_2[{\rm U}^{\rm BF}(6)]
\nonumber\\
&=&
2\left [\,
\hat U_1^{1/2} + \hat U_1^{3/2} + \hat U_2^{3/2} + \hat U_2^{5/2}
+ \hat U_3^{5/2} + \hat U_3^{7/2} \,\right ] ~,
\label{C-U6}
\ea
where
$\hat U_{\cal L}^{\cal J}\equiv\sqrt{2{\cal J}+1}
({\cal U}^{\cal L(J)}_+\tilde{\cal U}^{\cal L(J)}_-)^{(0)}_0$, 
and the quadratic Casimir of ${\rm SO}^{\rm BF}(6)$ is obtained for
\ba
(\hat{N} + \hat{n})(\hat{N} + \hat{n} +4) - \hat C_2[{\rm SO}^{\rm BF}(6)]
= \hat{{\cal M}}_{6} +
12\left [\hat V_0^{0} + \hat V_0^{1/2}\right ] ~,
\label{C-O6}
\ea
where
$\hat V_{\cal L}^{\cal J}\equiv\sqrt{2{\cal J}+1}
({\cal V}^{\cal L(J)}_+\tilde{\cal V}^{\cal L(J)}_-)^{(0)}_0$. 
In general, $\hat{H}'$ of Eq.~(\ref{Hprime}) is not invariant under 
${\rm U}^{\rm BF}(6)$ nor ${\rm SO}^{\rm BF}(6)$, 
yet the relations in Eqs.~(\ref{e_anni1})-(\ref{e_anni2})
ensure that a specific band of states will remain solvable with good
${\rm U}^{\rm BF}(6)$ and ${\rm SO}^{\rm BF}(6)$ 
quantum numbers $[N+1]\langle N+1\rangle$.
The combined effect of adding $\hat{H}'$ to the 
DS Hamiltonian~(\ref{e_hamds}),
$\hat{H}_{\rm PDS} = \hat{H}_{\rm DS} + \hat{H}'$, 
gives rise to a rich variety of Hamiltonians with PDS, 
for which only selected states are solvable with good symmetry,
while other states are mixed. 

\section{A Case Study: SO$^{\rm\bf BF}$(6) PDS in $^{\bf 195}\bf Pt$}

The SO(6) limit of the interacting boson model~\cite{Arima79} 
is known to be of relevance for the even-even
platinum isotopes~\cite{Cizewski78}. 
Accordingly, the classification~(\ref{e_clas}) 
is proposed in the context of the IBFM 
to describe odd-mass isotopes of platinum
with the odd neutron in the orbits $3p_{1/2}$, $3p_{3/2}$, and $2f_{5/2}$,
which are dominant for these isotopes~\cite{Isacker84,Bijker85}. 
In the current application of PDS to $^{195}$Pt,
we take a restricted Hamiltonian which, in the notation of 
Eqs.~(\ref{C-U6})-(\ref{C-O6}), has the form
\begin{eqnarray}
\hat H_{\rm PDS} &=&
\hat H_{\rm DS}+
a_0\hat V_0^{1/2}+
a'_1(2\hat U_1^{1/2}-\hat U_1^{3/2})
\nonumber\\
&&
+ a_2(\hat U_2^{3/2}+\hat U_2^{5/2})+
a_3(\hat U_3^{5/2}+\hat U_3^{7/2}) ~,
\label{e_hampds}
\end{eqnarray}
and $N=6$. 
These interactions can be transcribed
as tensors with total pseudo-orbital $\tilde L$ and pseudo-spin $\tilde S$
coupled to zero total angular momentum.
In particular, the $a'_1$ term in Eq.~(\ref{e_hampds}) has
$\tilde L\!=\!\tilde S\!=\!1$,
while the $a_0,\, a_2$ and $a_3$ terms have $\tilde L\!=\!\tilde S\!=\!0$.
\begin{figure*}[t!]
\centering
\includegraphics[width=13.5cm]{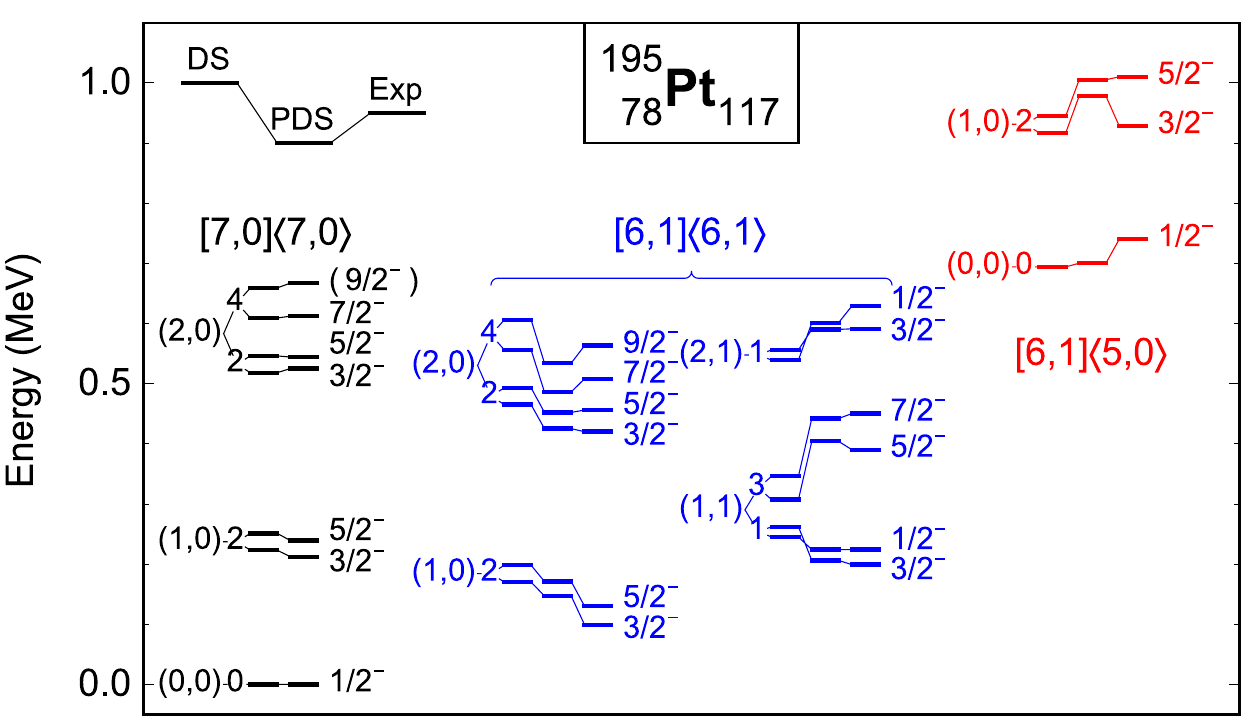}
\caption{(Color online). 
Observed and calculated energy spectrum of $^{195}$Pt.
The levels in black are the solvable $[7,0]\langle7,0\rangle$
eigenstates of $\hat H_{\rm DS}$~(\ref{e_hamds}),
whose structure and energy remain unaffected by the added PDS interactions
in Eq.~(\ref{e_hampds}).
The levels in blue (red) are the
$[6,1]\langle6,1\rangle$ ($[6,1]\langle5,0\rangle$) eigenstates
of $\hat H_{\rm DS}$~(\ref{e_hamds})
and are subsequently perturbed by the PDS interactions
in Eq.~(\ref{e_hampds}). Adapted from~\cite{isa15}.}
\label{f_pt195e}
\end{figure*}

\begin{figure*}[t!]
\centering
\includegraphics[width=8cm]{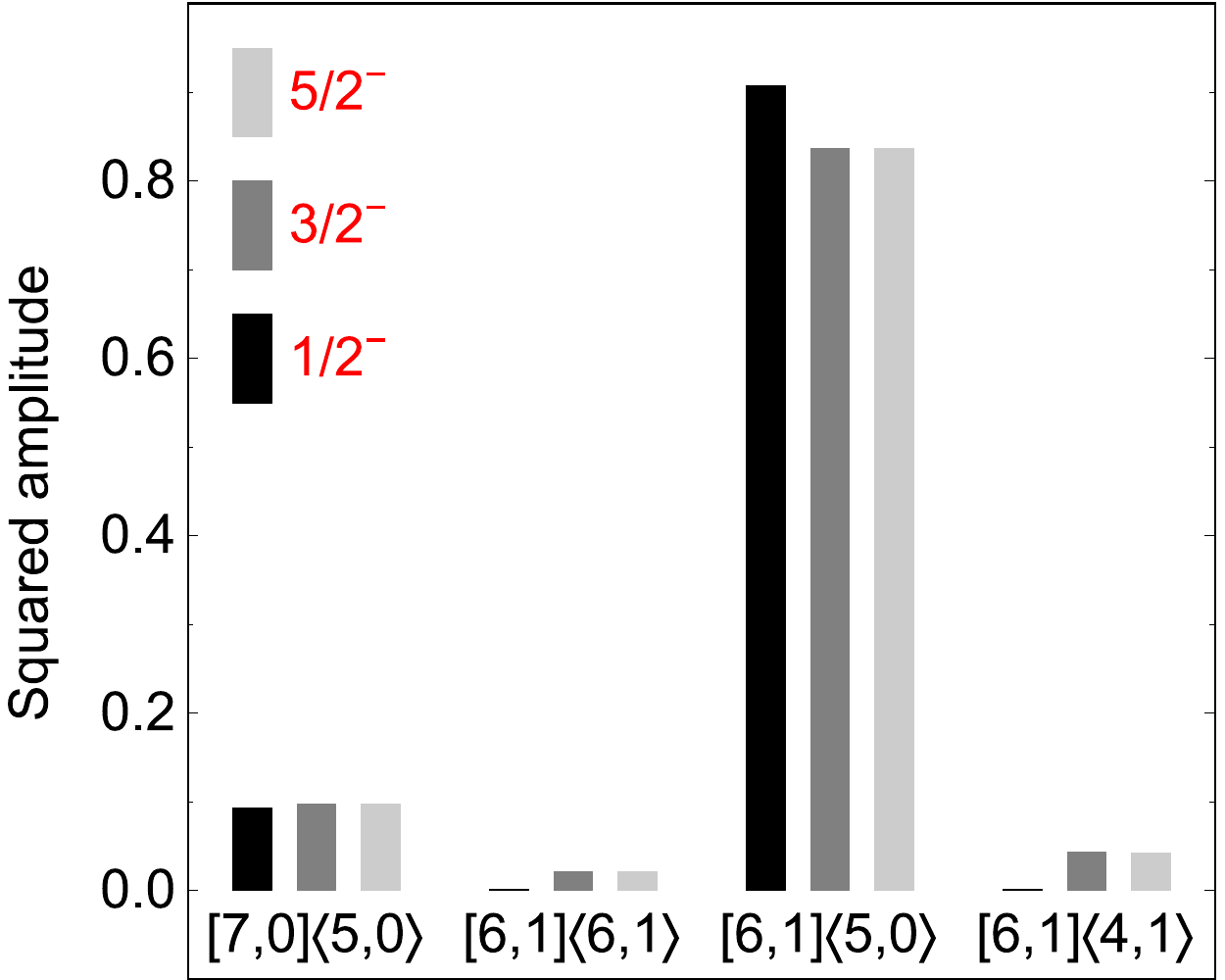}
\caption{(Color online). 
${\rm SO}^{\rm BF}(6)$ decomposition of the eigenstates of
$\hat H_{\rm PDS}$~(\ref{e_hampds}), shown in red 
in Fig.~\ref{f_pt195e}. Adapted from~\cite{isa15}.}
\label{f_ampli}
\end{figure*}

The experimental spectrum of $^{195}$Pt is shown in Fig.~\ref{f_pt195e},
compared with the DS and PDS calculations. The coefficients $c$, $d$,
and $d'$ in $\hat H_{\rm DS}$~(\ref{e_hamds}) are adjusted to the
excitation energies of the $[7,0]\langle7,0\rangle$ levels
which are reproduced with a root-mean-square (rms) deviation of 12~keV.
The remaining two coefficients $a$ and $b$ are obtained from an overall fit.
The resulting (DS) values are (in keV):
$a=45.3$, $b=-41.5$, $c=49.1$, $d=1.7$, and $d'=5.6$.
The fit for the PDS calculation proceeds in stages.
First, the parameters $c$, $d$, and $d'$ in Eq.~(\ref{e_hamds}) are taken
at their DS values.
This ensures the same spectrum for the $[7,0]\langle7,0\rangle$ levels
(drawn in black in Fig.~\ref{f_pt195e})
which remain eigenstates of $\hat H_{\rm PDS}$~(\ref{e_hampds}).
Next, one considers the $[6,1]\langle6,1\rangle$ levels
and improves their description by adding the three PDS $U$ interactions.
The resulting coefficients are (in keV):
$a'_1=10$, $a_2=-97$, and $a_3=112$.
Eq.~(\ref{e_anni1}) ensures that
the energies of the $[7,0]\langle7,0\rangle$ levels do not change
while the agreement for the $[6,1]\langle6,1\rangle$ levels is improved
(blue levels in Fig.~\ref{f_pt195e}).
The rms deviation for both classes of levels is 20~keV.
In particular, unlike in the DS calculation, it is possible to reproduce
the observed inversion of the $1/2^-$-$3/2^-$ doublets
without changing the order of other doublets.
The additional PDS terms necessitate a readjustment of the $a$ coefficient
in Eq.~(\ref{e_hamds}), for which the final (PDS) value is $a=37.7$~keV,
while the coefficient $b$ is kept unchanged.
Finally, the position of the $[6,1]\langle5,0\rangle$ levels
(red levels in Fig.~\ref{f_pt195e})
is corrected by considering the PDS $V$ interaction with $a_0=306$~keV
which, due to Eq.~(\ref{e_anni2}),
has a marginal effect on lower bands.
As seen in Fig.~\ref{f_pt195e}, the agreement is very good for yrast 
and non-yrast levels. 
As shown in Fig.~\ref{f_ampli},
while the states $[7,0]\langle7,0\rangle$ of the ground band are pure,
other eigenstates of $\hat{H}_{\rm PDS}$ in excited bands
can have substantial ${\rm SO}^{\rm BF}(6)$ mixing.

A large amount of information also exists
on electromagnetic transition rates and spectroscopic strengths.
In Table~\ref{t_be2}, 25 measured $B$(E2) values in $^{195}$Pt
are compared with the DS and PDS predictions.
The same E2 operator is used as in previous studies~\cite{Bruce85,Mauthofer86}
of the ${\rm SO}^{\rm BF}(6)$ limit,
$\hat{T}_\mu({\rm E2})=e_{\rm b}\hat{Q}^{\rm B}_\mu+e_{\rm f}\hat{Q}^{\rm F}_\mu$,
where $\hat{Q}^{\rm B}_\mu=s^\dag\tilde d_\mu+d^\dag_\mu s$
is the boson quadrupole operator,
$\hat{Q}^{\rm F}_\mu = K^{(2)}(2,0)+K^{(2)}(0,2)$ 
is its fermion analogue (see Table~\ref{Tab1}), 
and $e_{\rm b}$ and $e_{\rm f}$ are effective boson and fermion charges,
with the values $e_{\rm b}=-e_{\rm f}=0.151$~$e$b.
Table~\ref{t_be2} is subdivided in four parts
according to whether the initial and/or final state in the transition
has a DS structure (as in Refs.~\cite{Bruce85,Mauthofer86})
or whether it is mixed by the PDS interaction.
It is seen that when both have a DS structure
the $B$(E2) value does not change,
only slight differences occur when either the initial or the final state 
is mixed, and the biggest changes arise when both are mixed.
\begin{table}
\caption{\label{t_be2}
Observed $B({\rm E}2;J_{\rm i}\rightarrow J_{\rm f})$ values 
between negative-parity states in $^{195}$Pt
compared with the DS and PDS predictions of the ${\rm SO}^{\rm BF}(6)$ limit.
The solvable (mixed) states are members of the ground (excited) bands
shown in Fig.~\ref{f_pt195e}. 
Adapted from~\cite{isa15}.}\smallskip
\begin{small}\centering
\begin{tabular*}{\textwidth}{@{\extracolsep{\fill}}rrccrrccrrrrrr}
\hline  \noalign {\smallskip}
${E_{\rm i}}$&&$J_{\rm i}$&~~~&
${E_{\rm f}}$&&$J_{\rm f}$&~~~~~~&
\multicolumn{5}{c}
{$B({\rm E}2;J_{\rm i}\rightarrow J_{\rm f}$) ($10^{-3}~e^2{\rm b}^2$)}\\
\cline{9-13}
(keV)&&&&(keV)&&&&Exp&~~~~~~&DS&~~~~~~&PDS\\
\hline\noalign{\smallskip}
&\multicolumn{12}{c}{Solvable $\rightarrow$ solvable}&\\[1pt]
212&&${3/2}$&&0&&${1/2}$&&$190(10)$&&179&&179\\
239&&${5/2}$&&0&&${1/2}$&&$170(10)$&&179&&179\\
525&&${3/2}$&&0&&${1/2}$&&$17(1)$&&0&&0\\
525&&${3/2}$&&239&&${5/2}$&&$\le19$&&72&&72\\
544&&${5/2}$&&0&&${1/2}$&&$8(4)$&&0&&0\\
612&&${7/2}$&&212&&${3/2}$&&$170(70)$&&215&&215\\
667&&${9/2}$&&239&&${5/2}$&&$200(40)$&&239&&239\\[2pt]
&\multicolumn{12}{c}{Solvable $\rightarrow$ mixed}&\\[1pt]
239&&${5/2}$&&99&&${3/2}$&&$60(20)$&&0&&0\\
525&&${3/2}$&&99&&${3/2}$&&$\le33$&&7&&3\\
525&&${3/2}$&&130&&${5/2}$&&$9(5)$&&3&&2\\
612&&${7/2}$&&99&&${3/2}$&&$5(3)$&&9&&11\\
667&&${9/2}$&&130&&${5/2}$&&$12(3)$&&10&&12\\[2pt]
&\multicolumn{12}{c}{Mixed $\rightarrow$ solvable}&\\[1pt]
99&&${3/2}$&&0&&${1/2}$&&$38(6)$&&35&&34\\
130&&${5/2}$&&0&&${1/2}$&&$66(4)$&&35&&33\\
420&&${3/2}$&&0&&${1/2}$&&$15(1)$&&0&&0\\
456&&${5/2}$&&0&&${1/2}$&&$\le0.04$&&0&&0\\
508&&${7/2}$&&212&&${3/2}$&&$55(17)$&&20&&18\\
563&&${9/2}$&&239&&${5/2}$&&$91(22)$&&22&&22\\
199&&${3/2}$&&0&&${1/2}$&&$25(2)$&&0&&0\\
390&&${5/2}$&&0&&${1/2}$&&$7(1)$&&0&&0\\[2pt]
&\multicolumn{12}{c}{Mixed $\rightarrow$ mixed}&\\[1pt]
420&&${3/2}$&&99&&${3/2}$&&$5(4)$&&177&&165\\
508&&${7/2}$&&99&&${3/2}$&&$240(50)$&&228&&263\\
563&&${9/2}$&&130&&${5/2}$&&$240(40)$&&253&&284\\
390&&${5/2}$&&99&&${3/2}$&&$200(70)$&&219&&179\\
390&&${5/2}$&&130&&${5/2}$&&$\le14$&&55&&35\\
\hline
\end{tabular*}
\end{small}
\end{table}

\section{PDS and Intrinsic States}

An alternative way of constructing Hamiltonians with PDS for an algebra $G$,
is to identify $n$-particle operators which annihilate a lowest-weight
state of a prescribed $G$-irrep~\cite{Alhassid92}.
In the IBFM, such a state, which transforms as $[N+1]$ and $\tilde s=1/2$
under ${\rm U}^{\rm BF}(6)\otimes {\rm SU}^{\rm F}(2)$, is given~by
\begin{equation}
\vert \Psi_{\rm g}\rangle \propto
[b^{\dag}_c(\beta)]^Nf^{\dag}_{\tilde m_s}(\beta)\vert 0\rangle ~,
\label{Psi-g}
\end{equation}
where $b^{\dag}_{c}(\beta)  \propto (\beta\, d^{\dag}_0 + s^{\dag})$
and $f^{\dag}_{\tilde m_s}(\beta) \propto
(\beta\, c^{\dag}_{2,0;1/2,\tilde m_s} + c^{\dag}_{0,0;1/2,\tilde m_s})$
in the $\tilde\ell$-$\tilde s$ basis defined above.
$\vert \Psi_{\rm g}\rangle$ is a condensate of $N$ bosons and a single fermion,
and represents an intrinsic state for the ground band with 
deformation $\beta$. The Hermitian conjugate of the following
two-particle operators
\begin{subequations}
\begin{eqnarray}
{\cal V}^{0(0)}_+ &\propto&
{\textstyle\sqrt{5}}(d^\dag d^\dag)^{(0)}_0
-{\textstyle\beta^2}(s^\dag s^\dag)^{(0)}_0,
\label{Tb1}\\
{\cal V}^{0(1/2)}_{+,\mu} &\propto&
\sqrt{5}(d^{\dag} c^{\dag}_{2;1/2})^{0(1/2)}_{\mu}
-\beta^2(s^{\dag} c^{\dag}_{0;1/2})^{0(1/2)}_{\mu},\qquad
\label{Tb2}\\
{\cal U}^{{\cal L}({\cal J})}_{+,\mu} &\propto&
(d^{\dag} c^{\dag}_{2;1/2})^{{\cal L}({\cal J})}_{\mu},\quad
{\textstyle{\cal L}=1,3},
\label{Tb3}\\
{\cal U}^{2({\cal J})}_{+,\mu} &\propto&
(s^{\dag} c^{\dag}_{2;1/2})^{2({\cal J})}_{\mu}
- (d^{\dag} c^{\dag}_{0;1/2})^{2({\cal J})}_{\mu},
\label{Tb4}
\end{eqnarray}
\label{Tb}
\end{subequations}
satisfy $\tilde{\cal T}^{{\cal L}({\cal J})}_{-,\mu}\vert \Psi_{\rm g}\rangle=0$.
The $\cal V$ operators of Eqs.~(\ref{Tb1})-(\ref{Tb2})
satisfy also 
$\tilde{\cal V}^{0({\cal J})}_{-,\mu}\vert \Psi_{\rm e}\rangle=0$,
where
\begin{equation}
\vert \Psi_{\rm e}\rangle \propto
[b^{\dag}_c(\beta)\,c^{\dag}_{2,1;1/2,\tilde m_s}
- d^{\dag}_{1}\,f^{\dag}_{\tilde m_s}(\beta)][b^{\dag}_c(\beta)]^{N-1}
\vert 0\rangle
\label{Psi-e}
\end{equation}
is an intrinsic state, with ${\rm U}^{\rm BF}(6)$ label $[N,1]$,
representing an excited band in the odd-mass nucleus.
For $\beta=1$, $\vert \Psi_{\rm g}\rangle$ and $\vert \Psi_{\rm e}\rangle$
become the lowest-weight states in the ${\rm SO}^{\rm BF}(6)$ irreps
$\langle N+1\rangle$ and $\langle N,1\rangle$, respectively,
from which the $|(\tau_1,\tau_2)LJM_J\rangle$ states of Eq.~(\ref{e_anni2})
can be obtained by ${\rm SO}^{\rm BF}(5)$ projection,
and the operators~(\ref{Tb}) coincide with those listed in
Table~\ref{t_tensors}.

In case of axially-symmetric shapes, 
SO$^{\rm BF}$(5) is no longer a conserved symmetry and 
the following additional operators can contribute to 
Hamiltonians with other types of PDS,
\bsub
\ba
{\cal W}^{{2}(2)}_{+,\mu} &\propto&
\sqrt{2}\beta s^{\dag} d^{\dag}_{\mu}
+\sqrt{7} (d^{\dag} d^{\dag})^{(2)}_{\mu},\qquad\\
{\cal W}^{{2}({\cal J})}_{+,\mu} &\propto&
\beta (s^{\dag} c^{\dag}_{2;1/2})^{2({\cal J})}_{\mu}
\!+\! \beta (d^{\dag} c^{\dag}_{0;1/2})^{2({\cal J})}_{\mu}
\!+\!\sqrt{14} (d^{\dag} c^{\dag}_{2;1/2})^{2({\cal J})}_{\mu}.\;\qquad
\ea
\label{Tbaxial}
\esub
The above operators 
contain a mixture of components with different 
${\rm SO}^{\rm BF}(5)$ character ($\tau=1,2$), 
and annihilate the intrinsic states of 
Eqs.~(\ref{Psi-g}) and (\ref{Psi-e}). The solvable states are 
now obtained by angular momentum Spin$^{\rm BF}$(3) projection. 
The operators in Eqs.~(\ref{Tb}) and (\ref{Tbaxial}) 
are the Bose-Fermi analog of the 
proton-neutron boson-pair operators 
comprising the intrinsic part of the IBM-2 Hamiltonian~\cite{levkir90}, 
and used in the study of F-spin PDS~\cite{levgino00}.

\section{Summary and Outlook}

We have considered an extension of the PDS notion
to Bose-Fermi systems and exemplified it in $^{195}$Pt.
The analysis highlights the ability of a PDS to select
and add to the Hamiltonian, in a controlled fashion,
required symmetry-breaking terms,
yet retain a good DS for a segment of the spectrum.
These virtues greatly enhance the scope of
applications of algebraic modeling of complex systems.
The operators of Eqs.~(\ref{Tb}) and (\ref{Tbaxial}) 
can be used to explore additional PDSs in odd-mass nuclei, 
{\it e.g.}, SU$^{\rm BF}$(3) PDS for $\beta=\sqrt{2}$. 
Partial supersymmetry, of relevance to nuclei~\cite{Metz99}, can be studied
by embedding ${\rm U}^{\rm B}(6)\otimes{\rm U}^{\rm F}(12)$ 
in a graded Lie algebra.
Work in these directions is in progress.

\section*{ACKNOWLEDGMENTS}

The work reported was done in collaboration with P. Van Isacker (GANIL),
J.~Jolie and T.~Thomas (Cologne), and is supported
by the Israel Science Foundation.

\end{document}